\documentclass[12pt,preprint]{aastex}
\begin{document}
\normalsize

\shorttitle{Interacting Galactic Neutral Hydrogen Filaments }
\shortauthors{Verschuur}

\title{Interacting Galactic Neutral Hydrogen Filaments and Associated High-Frequency Continuum Emission}

\author{Gerrit L. Verschuur }

\affil{Physics Department, University of Memphis, Memphis, TN 38152}

\email{verschuur@aol.com}

\begin{abstract}
Galactic HI emission profiles in an area where several large-scale filaments at velocities  ranging from $-$46 km s$^{-1}$ to 0 km s$^{-1}$ overlap were decomposed into Gaussian components.  Eighteen families of components defined by similarities of center velocity and line width were identified and related to small-scale structure in the high-frequency continuum emission observed by the {\it WMAP} spacecraft, as evidenced in the {\it Internal Linear Combination} ({\it ILC}) map of Hinshaw et al. (2007).  When the center velocities of the Gaussian families, which summarize the properties of all the HI along the lines-of-sight in a given area, are used to focus on HI channel maps the phenomenon of close associations between HI and {\it ILC} peaks reported in previous papers is dramatically highlighted.  Of particular interest, each of two pairs of HI peaks straddles a continuum peak.  The previously hypothesized model for producing the continuum radiation (Verschuur, 2010) involving free-free emission from electrons is re-examined in the light of the new data.  By choosing reasonable values for the parameters required to evaluate the model, the distance for associated HI$-${\it ILC} features is of order 30 to 100 pc.  No associated H$\alpha$ radiation is expected because the electrons involved exist throughout the Milky Way.  The mechanism for clumping and separation of neutrals and electrons needs to be explored. 
\end{abstract}

\keywords{ ISM:atoms - ISM:clouds - cosmology}

 \section{Introduction}
In three previous reports (Verschuur, 2007a: Paper 1;  2007b; 2010: Paper 2) 108 close associations between foreground galactic neutral hydrogen (HI) features and small-scale structures found in the {\it Wilkinson Microwave Anisotropy Probe} ({\it WMAP}) {\it Internal Linear Combination} ({\it ILC}) map of Hinshaw et al. (2007) were listed.  The latter structures are purported to have a cosmic origin while the former are features within the disk of the Milky Way relatively close to the Sun.  In Paper 2 it was argued that a possible mechanism for generating the necessary high frequency continuum radiation is free-free emission from electrons produced by localized ionization of hydrogen.  However, no associated H$\alpha$ radiation is found and hence the model appears to fail. In this paper the mechanism is re-considered in the light of what is revealed of the HI properties through the Gaussian analysis of emission profiles.  

Since the onset of work on the associations between HI features and peaks in the {\it ILC} data, the manner in which the HI data have been considered has evolved.  In Paper 1, associations were found by comparing {\it ILC} data with structure found in HI area maps produced at 10 km s$^{-1}$ intervals, each covering an effective velocity range of 11.3 km s$^{-1}$.  In Paper 2 HI data were displayed in velocity maps made with a 3.3 km s$^{-1}$ effective bandwidth plotted every 2 km s$^{-1}$ in velocity.  These revealed the presence of the HI $-$ {\it ILC} associations more clearly.  The present study shows that when maps of the column density of HI Gaussian component families are compared to {\it ILC} features, the relationship becomes even more revealing.

In \S2 some of  the HI data used in this study are displayed and in \S3 the Gaussian analysis is described.  In \S4 the results of sorting Gaussian components into families are presented and their morphological relationship to the high-frequency continuum emission structure is shown in \S5.  The apparent failure of the previously suggested model for producing the high-frequency continuum radiation is considered in  \S6.  The discussion section in \S7 shows that the model does work when it is recognized that the necessary electrons already exist in the interstellar medium.  Conclusions are offered in \S8

\section{The Data}
The HI spectral line data used in this paper were drawn from the Leiden-Argentina-Bonn (LAB) All-Sky HI Survey (Kalberla et al. 2005) performed with a 0.\arcdeg6\ beam width and a 1.3 km s$^{-1}$ velocity resolution. 

To describe the context of the work to follow, Figure 1 displays five HI contour maps of brightness temperature in a 3.3 km s$^{-1}$ bandwidth as a function of galactic coordinates, longitude (GLON or {\it l}) and latitude (GLAT or {\it b}), for an area bordered by {\it l} $=$ 70\arcdeg\ \& 110\arcdeg\ and {\it b} $=$ 50\arcdeg\ \& 70\arcdeg.  Such maps were produced and examined at 2 km s$^{-1}$ intervals from $+$14 km s$^{-1}$ to $-$70 and then every 4 km s$^{-1}$ to $-$160 km s$^{-1}$ with respect to the local standard of rest.  Contour levels are shown in the captions.  The velocities of these displayed examples were chosen to illustrate several key points in the analysis to follow. 

Fig. 1a shows the HI brightness centered at a velocity of $-$46 km s$^{-1}$.   A curving filamentary feature can be followed from ({\it l,b}) $=$ (100\arcdeg, 70\arcdeg) to ({\it l,b}) $=$ (78\arcdeg, 50\arcdeg).  It will be referred to as Filament Alpha.  A weak extension from ({\it l,b}) $=$ (85\arcdeg, 57\arcdeg) to ({\it l,b}) $=$ (71\arcdeg, 50\arcdeg) suggests that filaments may be overlapping in the area around ({\it l,b}) $=$ (87\arcdeg, 59\arcdeg).  In the data of Hartmann \& Burton (1997: page 131) Filament Alpha can be followed well beyond the bounds of Fig. 1a.  Fig. 1b shows the HI morphology at $-$36 km s$^{-1}$. The HI distribution in the center is dominated by two peaks to be called South Pair.  They appear to lie on Filament Alpha.  However, other structure suggests that the picture may be more complex.

Fig. 1c shows HI brightness at $-$30 km s$^{-1}$.  A part of a twisted S$-$shaped filament, named Gamma, runs across the center of the area terminating in a bright feature at ({\it l,b}) $=$ (103\arcdeg, 61\arcdeg) before turning south.  The elongated peak at ({\it l,b}) $=$ (89\arcdeg, 56\arcdeg) appears to be an extension to this velocity of the southern component of South Pair.  Fig. 1d shows the HI brightness at $-$13 km s$^{-1}$ and a pair of peaks at ({\it l,b}) $=$ (91\arcdeg, 62\arcdeg)  lies on Filament Delta that stretches to the lower-right corner of the area.  These two peaks will be referred to as North Pair and they are located where Filament Delta crosses Filament Gamma.

Fig. 1e shows the HI brightness at $+$1 km s$^{-1}$.  The bright HI feature at ({\it l,b}) $=$ (87\arcdeg, 60\arcdeg)  lies on a complex filamentary feature labeled Beta and is located where Filament Beta overlaps Alpha; that is, between the two components of South Pair seen in Fig. 1b.  

Fig. 1f shows the positive amplitude {\it ILC} peaks and the two areas labeled 1 \& 2 marked by dashed lines are the boundaries within which Gaussian fitting was carried out.  The first-year {\it ILC} data used here is the same set used in Papers 1 \& 2 drawn from Hinshaw et al. (2007), effective beam width 1\arcdeg, mapped as contours from $+$0.02 mK in steps of 0.03 mK.  A visual comparison of the {\it ILC} data with this sample of HI maps shows associations between the two types of features.  A concentration of low-velocity HI in the region of enhanced {\it ILC} emission seen around  ({\it l,b}) $=$ (80\arcdeg, 57\arcdeg) is obvious.  To avoid the complexity manifest in that region, the present study was limited to the Areas 1 \& 2.  

Fig. 2 shows channel maps for these two areas at the velocities indicated in the caption with the same {\it ILC} contours shown in Fig 1f overlain.  The velocities were chosen based on the average velocities of individual Gaussian component families to be discussed below.  Fig. 2a shows low-level HI features that precisely skirt the boundary of extended {\it ILC} structure.  Fig. 2b shows the HI features labeled South Pair straddling an {\it ILC} peak and Fig. 2e shows North Pair.  What is striking is that without exception the small-scale HI peaks are associated with and closely offset from equally small-scale {\it ILC} features.  This offset between the two types of radiation is what was reported in Papers 1 \& 2.  However, only one of the associations reported in Paper 1, Source 16 at ({\it l,b}) $=$ (101.\arcdeg8, 59.\arcdeg5),  is seen here.  Its HI signature is evident in Figs. 2e \& f.   In Paper 1 the center velocity of S16 was listed as $-$5 km s$^{-1}$, but that analysis used area maps made every 10 km s$^{-1}$ covering a band width of 11.3 km s$^{-1}$.  Fig. 2f shows its peak to be at -9 km s$^{-1}$ and illustrates that more is learned about HI$-${\it ILC} associations when these narrower channel maps are used.  While Paper 1 only revealed one pair of associations in this area, because it only considered the brightest {\it ILC} peaks, the maps in Fig. 2 reveal at least nine more closely spaced associations with no cases of direct overlap, which is true for most of the 108 pairs listed in Papers 1 \& 2..    

At this point the analysis could end because Fig. 2 clearly confirms what was found in other sections of sky as outlined in Papers 1 \& 2.  In general small-scale structure in the high-frequency background continuum radiation maps observed by {\it WMAP} as revealed in the {\it ILC} data is associated with, but slightly offset from, structure in galactic HI emission..  However, noting associations reveals nothing substantive about the physical properties of the HI being observed; for example, what are the volume densities and temperatures of the HI and what process could give rise to the associated high-frequency continuum emission?   As the next step in this process, the efficacy of using Gaussian analysis to understand the HI properties was explored to determine if this clarifies the underlying physics that might, in turn, help understand the nature of the emission process giving rise to the high-frequency continuum radiation.

\section{Gaussian Analysis}
The HI area maps discussed so far reveal only the amplitude of the HI emission at a given velocity; they reveal little about the physics or dynamics of the structures that are present.  HI emission profiles usually consist of multiple Gaussian components produced by concentrations of HI along the line-of-sight or by motion or temperature variations within coherent sub-features within a given volume of space. Each Gaussian is characterized by a peak brightness temperature, $T_B$(max), a center velocity, $v_c$, and a line width, {\it W}, the full width at half maximum brightness.  The area maps in Figs. 1 \& 2 represent the total of the brightness temperatures at any given velocity that is produced by overlapping components. The goal of the present study translates to untangling the Gaussian structure to identify the properties of the overlapping components.  

The hypothesis that the HI component line shape is Gaussian rests on the assumption that motions within any mass of HI will show a random velocity distribution that can be described by a Gaussian function.  However, Gaussian analysis of HI profiles is difficult because a given solution may not be unique and that means that relatively few researchers have ventured into the realm of such analyses.  The challenge has been discussed by Verschuur (2004), Verschuur \& Peratt (1999), and Verschuur \& Schmelz (2010).  The algorithm used in the present Gaussian decomposition of LAB profiles was described by Verschuur (2004).  A crucial difference between this approach and the totally automated analysis used by, for example, Haud (2000) and subsequently by Haud \& Kalberla (2007), is that in the present study each profile and the associated Gaussian fit is examined visually to assure that the results are consistent.  The problem is that the Gauss fitting algorithm is set to minimize residuals and false minima can be found for a set of Gaussian components that bear no relationship to the profiles in their immediate vicinity.  While other researchers may have attempted to design their algorithms to take in account neighboring solutions, in our experience that is not necessarily sufficient, as will illustrated with examples to follow.

Gaussian decomposition of HI profiles in the present study was performed for Areas 1 \& 2 (Fig. 1f) using profiles every 1\arcdeg\ in {\it l} and 0.\arcdeg5 in {\it b}.  At the latitude of these areas (60\arcdeg) this created a uniform sampling grid of 588 profiles separated in real angle of 0.\arcdeg5 in both coordinates.  The Gaussian decompositions allowed the fitting of up to 9 components although very few profiles actually required that many.  

During the analysis it became clear that Gaussian decomposition is subject not only to the presence of noise but is especially vulnerable to non-noise-like low-level (interference?) features found even in the baseline areas of the LAB profiles where no HI is present.  In some profiles these showed a negative amplitude.  When they occured at a velocity at the edge of a profile, the properties of the component fit that included the relevant velocity range could be ÒdraggedÓ to apparent line widths that were either higher or lower than those found for similar components in neighboring profiles. This problem could only be identified by visually examining the Gaussian fits.  Obviously there is no way to detect the presence of such non-noise-like deviations within the velocity range of the HI emission itself, which contributes to noise in the values of the derived component parameters.  Visual examination of each Gauss fit compared to neighboring profiles allows the distortion produced by these non-noise structures near the profile edges to be recognized and the algorithm was then run with a different initial setting to produce a results consistent with neighboring profiles.  In several profiles small interference spikes in only one channel were present and in two profiles single-channel negative spikes that could be confused with narrow absorption lines occurred at velocities that also contributed to producing spurious Gaussian parameters.  These obvious cases were dealt with by canceling the problem spike in a given channel.  It is highly doubtful that an automatic Gaussian analysis could deal with these problems unless it used advanced AI techniques.

To initiate the search for the best-fit Gaussians for any given profile, the results for an adjacent solution were used.  When the results of a given initialization were clearly inconsistent with the trends observed in surrounding profiles, the profile was re-initialized using another of the adjacent Gauss fits.  Note that while the process of fitting Gaussian components to individual profiles is inherently noisy, the data for an ensemble of fits do produce coherent patterns in column density maps that allow families of components to be recognized and mapped.  

A pervasive, underlying broad line width component is the signature of every profile, both at LV and IV.  Its width is predominantly about 33 km s$^{-1}$.  However, after the initial analysis and sorting of families of Gaussian and making maps of the HI column densities in Area 2, a systematic error was recognized that was readily rectified.  The data showed that the broad components often manifested line widths of order 24 km s$^{-1}$ in addition to the 33 km s$^{-1}$ regime.  When maps of the narrow line width components were made it was found that the dominant one, called LV1, to be discussed below, was absent in the those directions where 24 km s$^{-1}$ wide components were found.  Thus the initially produced area map of the HI column density for LV1 had ÒholesÓ in the directions where the Gauss fitting routine had found a solution involving  a broad component of order 24 km s$^{-1}$ wide.

The profiles in those direction were therefore reconsidered and it was found that a significantly better fit could be obtained by initializing the algorithm using the nearest solutions in which the 33 km s$^{-1}$ lines was present.  Thus the 24 km s$^{-1}$ components turned out to be the result of a self-perpetuating fit that produced false minima.  The relevant profiles were rerun and subsequent mapping of component LV1, using the better Gauss fits, did not show any holes.  In fact LV1 is found to be pervasive, as are the underlying broad components of order 33 km s$^{-1}$ wide, see tables below for details.  This unfortunate state of affairs and its rectification is highlighted because automated Gaussian fitting programs are subject to the problem of false minima in residual phase space and future attempts to duplicate the present study should pay close attention to the result for given profiles to make sure no such systematic effects creep in.  Fortunately, such effects stamp their presence of otherwise ordered column density maps, which allows their presence to be recognized.

Figure 3 illustrates several examples of Gaussian fitting to a variety of profiles at the positions indicated.  Fig. 3a shows a profile defined by a single, underlying broad component with line width {\it W} $=$ 36.2 km s$^{-1}$ at ${v_c} = -$8.9 km s$^{-1}$.  The profile just 1\arcdeg\ away in longitude shown in Fig. 3b reveals the presence of a second underlying broad component at intermediate velocity (IV) ${v_c} = -$30.0 km s$^{-1}$ with {\it W} $=$ 35.1 km s$^{-1}$ while the other is at low velocity (LV) centered at ${v_c} = +$0.7 km s$^{-1}$ with {\it W} $=$ 33.9 km s$^{-1}$.  The separation on the sky of these two profiles is 0.\arcdeg5\ in true angle and serves to illustrate the tremendous amount of structure found in the HI distribution.  Fig. 3c shows two broad underlying components whose presence is easy to discern because they manifest so clearly in the wings of the overall profile.  The broad IV component has {\it W} $=$ 35.1 km s$^{-1}$ while the LV broad component has {\it W} $=$ 34.8 km s$^{-1}$.  

The profile displayed in Fig. 3d is notable because of the presence of a narrow component at the center of the profile, only 3.4 km s$^{-1}$ wide. The underlying broad components for this profile are 37.5 km s$^{-1}$ wide at IV and 37.2 km s$^{-1}$ wide at LV.  In adjacent profiles the amplitude of the HI emission between velocities of $-$80 and $-$30 km s$^{-1}$ produced an almost straight line slope and solutions were only obtained by painstakingly using the surrounding profiles to find a fit consistent with the trends in the data around those positions.  It seems very unlikely that an automated program would find solutions to such a profile.  Fig. 3e shows an extremely bright and narrow center component 3.0 km s$^{-1}$ wide at $-$17.2 km s$^{-1}$, which is part of component family LV13, see below.  The associated broad components are 27.8 km s$^{-1}$ wide at $-$47.8 km s$^{-1}$ and 36.8 km s$^{-1}$ wide at $-$5.0 km s$^{-1}$.

Fig. 3f shows a profile where the peak brightness temperature at IV and LV are maxima for the entire mapped area.  The underlying broad components have {\it W} $=$ 30.6 km s$^{-1}$ at IV and {\it W} $=$ 32.0 km s$^{-1}$ at LV.  The fact that both peaks reach maxima here is an indication that the structures at IV and LV are directly related, as will be discussed in the context of examining all the data, below.

\section{Sorting Gaussians into parameter families }
Gaussian analysis of 288 profiles yielded 3,706 Gaussian components that were sorted into 18 distinct families most of which were recognized in both Areas 1 \&2. The sorting of Gaussians into families was usually straightforward because of clear continuity in center velocity and/or line width that could be followed across many adjacent profiles.  When the data were gathered into families and plotted as area maps, the presence of components that had not been assigned to the correct family was quickly apparent.  They could then be assigned to the appropriate family and the resulting maps show a high degree of order as will be displayed below.  Only six components could not be fit into any of the families and may be regarded as noise in the data.

Table 1 summarizes the key properties of component families for the Area 1 and Table 2 applies to Area 2.  Col. 1 gives the name of the family, Col. 2  lists the number of Gaussians used to define the family, Col. 3 lists the average center velocity, $\overline{v_c}$, for the family and Col. 4 its average line width (full-width, half-maximum), $\overline{\it W}$, both with one standard deviation errors.  Col. 5 lists the peak column density found for a given family.  Table 3 summarizes all the family data.  Col. 1 is again the family name, Col. 2 is the total HI column density summed over all the members of that family in both Areas and Col. 3 is the fraction of the total column density for a given family.  The average column density of the family is given in Col. 4 and the peak value in Col.5, which corresponds to the value in either Table 1 or 2.  Of these component families three (Broad LV, Broad IV and LV1) were found in all directions, although for some of the Broad IV features the amplitudes are so low as to be comparable to the peak-to-peak noise in the data.  

For all the broad components listed in Tables 1 \& 2, which includes 1176 cases,  $\overline{\it W} =$ 33.1$\pm$1.1 km s$^{-1}$.  A simple test was performed to show that this large line width is not an artifact of the observing beam width encompassing velocity gradients within the beam to produce the higher line width values.  HI profiles obtained by Verschuur  (2013) toward a high-velocity HI feature known as A0 using the Green Bank Telescope (9.\arcmin1 beam width) were Gaussian analyzed and the results compared with a similar decomposition of profiles from the LAB survey toward the same feature. Both sets of data revealed an underlying broad component about 22 km s$^{-1}$ wide with no hint of a 33 km s$^{-1}$ wide component.  Thus, in general, it is not simply an artifact of lower angular resolution that generates the 33 km s$^{-1}$ wide components.  It is notable that the present study did not reveal any convincing evidence for components that could be associated with a so-called WNM, of Warm Neutral Medium, at 8,000 K, which would produce line width of order 20 km s$^{-1}$.  However, the data in Table 2 show that for two component families labeled IV-HV and HV the average line widths are in the 20 km s$^{-1}$ regime.  That is a property of anomalous velocity HI, not the gas at lower velocities.   [In an unrelated and as yet unpublished study of the Gaussian component structure of high-velocity clouds by the author, the LAB data show that a component of order 22 km s$^{-1}$ wide is common and also that another component about 34 km s$^{-1}$ wide is present in many cases.  In some directions, the HI emission profile can be fit by only one Gaussian, which is then found to be either about 22 or 34 km s$^{-1}$ wide.] 

When compared to Gaussian line width data obtained from totally unrelated studies the numerical value for the broad component line width is striking.  Verschuur \& Schmelz (2010) gathered together line width data in their Table 2 pertaining to an underlying broad line-width ÒbackgroundÓ component reported in 9 published papers by various researchers as well as their own data that together produced $\overline{W} =$ 33.7$\pm$2.4 km s$^{-1}$.  This is essentially identical to the value of 33.1$\pm$1.1 km s$^{-1}$ in the present study.  [The possible role of a mysterious plasma phenomenon known as the Critical Ionization Velocity effect in affecting the line widths of the broad components is discussed by Verschuur (2007b) and Verschuur \& Schmelz (2010) and references therein.  Further consideration of that poorly understood phenomenon is beyond the scope of the present study.]  Several other families listed in Tables 1 \& 2 include 1139 components with a weighted average line width of 14.6 $\pm$1.2 km s$^{-1}$.  Again, this value is significant when compared to the results summarized by Verschuur \& Schmelz (2010) in their Table 3 of 13.9$\pm$0.9 km s$^{-1}$.  For completeness, 10 entries in Tables 1 \& 2 have $\overline{W}$ values between 4.1 \& 8.1 km s$^{-1}$ involving 923 components for a weighted average line width of 5.7$\pm$0.6 km s$^{-1}$.  The remaining 339 components have average line widths between 8.7 \& 10.7 km s$^{-1}$ to produce a weighted average line width of 9.6$\pm$1.3 km s$^{-1}$.  

 \section{The relationship between the HI component families and {\it ILC} peaks}
The HI column density maps for the families of components are next plotted and compared with {\it ILC} data.  What will become apparent is that a great deal more remains to be revealed about the crucial HI structure in future higher resolution studies.

Figures 4a $-$ c show the morphology of the families of HI components at intermediate velocities from $-$48 to $-$35 km s$^{-1}$ that encompass the emission from South Pair at the right-hand side.  The {\it ILC} peak at {\it l,b} $=$ (88.\arcdeg5, 58.\arcdeg5) bridges the space between the major peaks in the HI defining South Pair.  A fourth IV component at $-$55 km s$^{-1}$ seen in Fig. 4d is associated with {\it ILC} peaks at the top-left of the area.  It is striking that South Pair is clearly defined by HI components with three distinct line widths, see Table 1.  The Gaussian analysis does not reveal the greater extent of Filament Alpha in these HI column density maps since column density is given by the product of line width and amplitude.  An intrinsically narrower yet bright feature may not stand out against a background of somewhat wider components, even if it is a member of the same family.  The presence of three distinct component families with different line widths at the location of the northern half of South Pair may be indicative of the contributions from three distinctly  different filaments whose existence can be inferred by close examination of Fig. 1.

Figure 5 shows area maps of the HI column densities for the families with $\overline{v_c}$ between  $-$26 \& $-$19 km s$^{-1}$, LV11, 12, 13 \& 14, compared to the{\it ILC} contours.  Fig. 5a shows the patchy morphology of LV11 and its brightest peak appears to be associated with North Pair and another {\it ILC} peak at the left-hand edge of the map around {\it l} $=$ 100\arcdeg\ is found in the region of more complex  {\it ILC} structure.  The members of the LV11 family were identified by similarities in line width.   Fig. 5b shows the morphology of LV12 that is clearly related to the presence of the {\it ILC} peaks.  Its average line width is similar to LV11 but its center velocities are very different, Table 1.   Fig. 5c shows LV13 \& 14 combined since their morphologies smoothly blend although their velocities show a considerable change from Area 1 to Area 2, see tables.  Again, HI column density peaks are clearly associated with {\it ILC} peaks.  When combined in Fig. 5d these four HI component families are located where several filaments intersect or overlap, see  Figs. 1c, d \&e and the overall feature in Fig. 5d thus manifests variations along its length of center velocities and line widths.  Together these patterns imply that a great deal of complexity is hidden from view because of the poor angular resolution of the available data. This is borne out by the HI profiles in the region of North Pair.  They are extremely complex, see, for example, Fig. 3d, which makes the task of confidently identifying families of line widths in this area very difficult.  What remains true, however, is that the HI column density maps reinforce the notion that HI and {\it ILC} structures are related, as is so clearly evident in Fig. 2.  

Fig. 6 shows the morphology of the low velocity components.  Fig. 6a plots the {\it ILC} contours on a map of component family LV1 that is found throughout the area and is therefore a pervasive background or $"$field$"$ component.  Its morphology is not obviously related to the presence of {\it ILC} structure.  Fig. 6b shows the morphology of LV2, which is part of Filament Beta (Fig. 1e) and this segment of filamentary structure is clearly related to the presence of the {\it ILC} peak where it appears to terminate at ({\it l,b} $=$ 88\arcdeg, 58\arcdeg) but examination of  Fig. 1e suggests it then curves back and away to the west from that terminus.  

The high-frequency continuum radiation creating the {\it ILC} peak in South Pair originates in a volume of space where the two filaments, Alpha and Beta overlap.  Also, very striking, is the fact that where the HI peak in IV-A ($\overline{v_c} =  -$38.6 km s$^{-1}$) overlaps LV3 ($\overline{v_c} = -0.4$ km s$^{-1}$) at {\it l,b} $=$ (87.\arcdeg0, 59.\arcdeg5) the HI brightness of both the IV and LV component is by far the greatest for the area mapped.  The relevant profile is shown in Fig. 3f.  This is strong circumstantial evidence that HI at very different velocities is somehow interacting and hence at the same distance.  This phenomenon that HI at very different velocities is directly associated was also found in several areas described Paper 1.  Fig. 6c shows the morphology of LV3 and its properties are so similar to LV2 that they could well be combined, yet they appear to represent two distinct families in one filamentary feature.

The other three frames in Fig. 6 show the HI column density maps for the remaining LV component families and the fact that the majority of the structures, albeit patchy, seem to be associated with immediately adjacent {\it ILC} features suggests that there is more to be learned from higher-resolution HI mapping.  The HI associated with S16 discussed above can be seen in Figs. 6d \& e and in this display is less dramatic than found in Figs. 2e \& f.

For completeness, Fig. 7 gathers together the column density maps for the remaining component families.   Fig. 7a shows the HI morphology for the LV broad line-width component and Fig. 7b is the same for the IV broad line-width family.  There is some indication for diagonal features but these area maps are not readily related to the {\it ILC} structures.   An exception is the bright patch just to the south of the North Pair {\it ILC} feature in Fig. 7a.  It contributes to the pattern seen in Fig. 2b, center frame.  Examination of the HI profile in this direction again shows it to be extremely complex, similar to the profile in Fig. 3d, and the complexity can only be untangled using higher resolution HI data.  

The final two frames in Fig. 7 refer to component families that were not a carefully mapped because their brightness temperatures are very low, barely above peak-to-peak noise in some cases.  The case of Component IV-HV ($\overline{v_c} = - $78 km s$^{-1}$) is shown in Fig. 7c.  In some directions its spectrum is well separated from the bulk of the HI emission in velocity while in other directions it overlaps the wings of the HI emission from the bulk of the IV gas.  Yet, its peak column density lies toward the North Pair {\it ILC} peak.  Finally, Fig 7d shows the column density map for weak high-velocity HI, called Component HV ($\overline{v_c} =  -$109 km s$^{-1}$), which favors other peaks in the {\it ILC} map.  

\subsection{What the Gaussian mapping reveals}
Overall, the goal of mapping Gaussian components with a view to clarifying the relationship between HI and {\it ILC} peaks has done little to reveal a more edifying picture. Instead the picture has become more complex.   On the one hand there is strong evidence that HI at very different velocities appears to be involved where {\it ILC} peaks are located, and on the other hand those interactions are complex and somehow give rise to the small-scale structure in the high-frequency continuum radiation found in the {\it ILC} map.  The patterns seen in the HI area maps displayed in Fig. 2 are also found in the Component maps, Figs. 4 $-$ 7, so the area maps at specific velocities determined by Gaussian family average velocities may be the most effective way to identify close associations between HI and small-scale {\it ILC} structure.  This reduces the task from having to sort through 60 to 100 HI channel maps to find associations to considering only a dozen or so data sets determined by the identification of Gaussian families.  Bear in mind that the HI column density maps of the ensemble of component families contain between them information about ALL the HI found over the total line-of-sight through any given area.

\subsection{A note on the statistics of associations}
In Paper 1 an attempt was made to show that the association between HI and slightly offset {\it ILC} features was significant.  In Paper 2 it was shown that there is no evidence for widespread direct positional associations, which was never claimed in any case.  Yet the question remained as to whether the near positional associations discussed here and in Papers 1 \& 2 are significant or due to chance.  

Interstellar HI structure is seen all over the sky and so is the structure revealed by {\it ILC} data.  However, for any given area of sky as many as 100 HI area maps at velocities separated by 2 km s$^{-1}$ need to be studied and the likelihood of chance associations becomes large. At the same time, the data discussed above and in Papers 1 \& 2 show that no two examples of associations are identical.  Providing convincing statistical arguments that a given association between an HI feature and one found in the {\it ILC} data is significant may be impossible.  There is no {\it a priori} standard for defining an association that can be tested for statistically.  Instead we must rely on looking at the actual data that show the associations clearly.  

Seen from another perspective, the entire sky is filled with small-scale structure found in the high-frequency continuum emission observed by {\it WMAP} as revealed in the {\it ILC} map, which was produced after possible sources of intervening radiation had been removed, Hinshaw at al. (2007).  But interstellar HI structure also covers the entire sky and that was not taken into account in the production of the {\it ILC} map.  There would have been no {\it a priori} reason for doing so.  While random coincidences in position between {\it ILC} and HI peaks are to be expected, this is not what is found.  In general, apparently associated {\it ILC} and HI features are offset from one another by a small amount, on average 0.\arcdeg8, as stated in Paper 2, and estimated to be closer to 1\arcdeg\ for the patterns seen in Fig. 2.  

\section{The apparent failure of the previous model}
The area maps presented in Fig. 2 reveal the presence of several sets of associated HI $-$ {\it ILC} features.  In Paper 2 it was suggested that the continuum radiation is produced by free-free emission from electrons interacting with electrons and that the excess electrons are produced by the ionization of hydrogen atoms where HI features interact. It was also noted that for the emission mechanism to account for the data, the continuum sources would have to be unresolved in the {\it WMAP} survey in order for the presence of the phenomenon to have been missed in the original {\it ILC} analysis of the {\it WMAP} data. 

For the present data an order-of-magnitude calculation can be performed.  For example for North Pair using the HI column density of the associated peak of 67 $\times$ 10$^{18}$ cm$^{-2}$ (Fig. 5d), it is possible to calculate (using the formula derived in Paper 2, see Eqtn. 1, below) the degree of ionization required to produce the observed {\it ILC} peak of 0.08 mK.  For a distance of 100 pc, an electron excitation temperature of 8,000 K, a source width of 0.\arcdeg2,  and a depth along the line of sight equal to that width, the fractional ionization required is 0.45.  The electron density that is implied is 28.3 cm$^{-3}$ and the corresponding emission measure (given that the path length is known for a given distance) should produce H$\alpha$ emission at a level of 125 R in the 1\arcdeg\ beam of the WHAM survey of Haffner et al. (2003).  However, examination of the WHAM data reveals no such structure.  The maximum emission measure over the relevant area of sky is 0.7 R, barely above a 0.5 R background level.  Therefore, the hypothesis that the continuum peaks in the {\it ILC} data are due to free-free emission from electrons produced through localized ionization of the HI in the galactic disk, as suggested in Paper 2, fails.  No combination of parameters (distance, angular size, aspect ratio, and excitation temperature) can account for both the observed continuum emission level and the lack of H$\alpha$ signal.  But what, then, is the source for the high-frequency continuum radiation if the associations seen in Fig. 2 are real?

\section{Discussion}
The original goal of this project was to determine the properties of two pairs of HI features in directions where filaments cross and at which locations peaks in the {\it ILC} data were found. As a result of mapping families of Gaussian components in the area, the HI distribution was found to be more complex than the impression garnered from examining up to 100 area maps of brightness temperature at specific velocities at 2 km s$^{-1}$, for example.  In instead the average velocities of the Gaussian families are used to focus attention on specific HI channel maps the relationship between HI structure and {\it ILC} structure emerges in dramatic detail, see Fig. 2.  In fact, Fig. 2 \& Figs. 4 $-$ 7 confirm the claims made in Papers 1 \& 2 that close associations between small-scale HI and {\it ILC} features are real.  However it is not immediately clear how the high-frequency continuum emission is generated in local interstellar space, local because the HI data discussed here is at high galactic latitudes of 60\arcdeg\  and the velocities imply a galactic origin.  As this discussion proceeds it will become increasingly obvious that we are venturing into uncharted territory.  The most fundamental issue that has yet to be resolved is at what angular scale both the HI and the {\it ILC} structures can be claimed to be resolved.  Given that uncertainty it will be shown below that it is nevertheless possible to obtain an apparently good fit to the data using the model proposed in Paper 2.

A solution requires that several questions be answered.  First, given that the free-free emission from electrons appears to work quite well as outlined in Paper 2, other than the absence of associated H$\alpha$ radiation, what might be the source of electrons so as to avoid the H$\alpha$ dilemma?  A second question concerns what mechanism is operating to cause the HI and the electrons to cluster?  Also, why would the neutrals and electrons cluster in slightly offset directions?  Lastly, if a source of electrons can be identified and a mechanism for clustering suggested, is it still possible to account for production of the high-frequency continuum radiation through invoking free-free emission from electrons?

\subsection{Source of electrons}
It is well known that free electrons exist everywhere in interstellar space as inferred from pulsar dispersion measures and radio source rotation measures.  Average electron densities along path lengths of hundreds to thousands of pc are estimated to be in the range 0.03 to 0.3 cm$^{-3}$ (see, for example, Wood \& Linsky, 1997; Allen, Snow \& Jenkins, 1990; Lyne, Manchester \& Taylor, 1985).  Unfortunately, essentially nothing is known about the clumping of these electrons along a given line-of-sight.  In contrast, the clumping of the HI is its most basic characteristic, producing morphologies such as seen in Fig. 1.  Using the data in Table 3 the average HI column density for the 588 lines-of-sight included in the study is 123 $\times$ 10$^{18}$ cm$^{-2}$. Assuming a typical path length of 100 pc the average HI volume density is 0.41 cm$^{-3}$.  An average electron density of 10\% of the HI, that is 0.04 cm$^{-3}$,  lies in the range interstellar electron density from pulsar dispersion measure data.  This offers a first-order approach to determining whether a reasonable set of parameters can be found that can be used with the electron Brehmstrahlung model to account for the observed amplitudes of the high-frequency continuum signals found in the {\it ILC} data.

\subsection{Application of the theory}
From Eqtn. 12 in Paper 2, the brightness temperature $T_B(\nu)$ of the high-frequency continuum radiation at a frequency $\nu$ produced by free-free emission from (cold) electrons with an excitation temperature, $T_e$, is given by: 
\begin{equation}
T_B(\nu) = 1.86\times10^{17}\  \nu^{-2}\ T^{-0.5}_e\ ln (4.7 \times 10^{10}\ T_e/\nu)\ \it f^2\ N_H^2\  (\theta_o\ A\ L)^{-1}\  K,
\end{equation}
where $N_H$\ is the observed HI column density in units of 10$^{18}$ cm$^{-2}$ and the electron density is expressed as a fractional degree of ionization, {\it  f}.  The angular width of the electron enhancements (or cloud) on the sky is $\theta$$_o$ and A is the Aspect Ratio, the depth of feature relative to its width.  L is the distance in pc. 

In order to evaluate Eqtn. 1, several assumptions have to be made and then, based on what is found, the direction of future research may be indicated.  The electron excitation temperature is set to 100 K, the typical kinetic temperature of interstellar neutral hydrogen, bearing in mind that the data show little evidence for a Warm Neutral Medium at 8,000 K, as noted in \S4.  The reasons for the Gaussian family line width much greater than expected for this temperature have been alluded to in \S4 and deserves a detailed discussion beyond the bounds of the present study. [Such a discussion would in any case conclude that electrons exist in abundance that are not created by localized ionization of neutral hydrogen.]  An angular width of the small-scale {\it ILC} features has to be assumed since the structures considered above are usually about 1\arcdeg\ across, which is the resolution of the {\it ILC} map.  Thus it is fair to assume that the sources of high-frequency continuum radiation are unresolved on this scale.  (Others with access to high resolution observations of the high-frequency continuum radiation should look into this issue; e.g., those who use the Planck spacecraft data.)   In order to explore whether Eqtn. 1 works, the model amplitudes are calculated for the two ends of the {\it WMAP} band, 23 \& 94 GHz, and then averaged.  

Eqtn. 1 is applied to the two cases of close associations between HI and {\it ILC} peaks, North Pair and South Pair, and it is used to determine the distance at which it can account for the observed {\it ILC} positive amplitudes as a function of the required degree of ionization of the associated HI column density (as a first-order approach to the data).  The results are shown in Fig. 9.  For example, if the angular scale of the unresolved {\it ILC} peak in South Pair is 0.\arcdeg1\ then the solid line in Fig. 9 so labeled indicates that for a range of electron densities equal to 0.10 to 0.17 times the associated HI peaks the distance of the source required to produce the observed {\it ILC} amplitude of 0.12 mK would be between about 30 and 100 pc.  The calculation used the peak HI column density for IV-A as the guide to the column density of the associated electron cloud.  It remains to be determined just what value should be used given that there are several distinct HI families of line widths involved in this direction, see Fig. 4.  But to first-order the model does match the data.  The difference between this calculation and the one reported in \S6 is that the electron temperature is not 8,000 K as would be expected from localized ionization of HI but closer to 100 K consistent with the temperature of the cold hydrogen atoms.

The fact that the curves in Fig. 9 for North Pair and South Pair appear to encompass a region of phase space that is reasonable as regards the required electron column densities and distances, for the 100 K regime, indicates that the possibility that the {\it ILC} structures are indeed located in the galactic disk relatively close to the Sun should be seriously considered.  But what mechanism would simultaneously act to clump the neutrals and the electrons and have them physically separated yet close associated in space.

\subsection{On the clumping and separation of electrons with respect to HI}
Fig. 1 shows clear evidence for the presence of several large-scale filaments of HI in the area under consideration and Fig. 2 shows that HI and {\it ILC} features over the target area for this study are connected and offset from one another.  These are observational facts that have to be recognized. There appear to be several ways in which electrons and neutrals could become spatially separated in interstellar space, although none has been formally studied for such an environment.  The options are only briefly mentioned here.

In a completely different astrophysical situation, namely the solar environment, many papers have deal with a phenomenon called the First Ionization Potential (FIP) Effect.  It is used to account for the spatial separation either in layers in the solar atmosphere, or along flux tubes, of various atomic species.  The FIP Effect is invoked to account for solar abundance variations that are otherwise difficult to comprehend. A number of references that indicate how the FIP Effect may be important include Raymond (1999), Laming (2009) and Schmelz et al. (2012).  Another way of considering this is to recognize that the offset between the HI and {\it ILC} peaks is one of e/H abundance variations in interstellar space, either along a given filament or between adjacent volumes of space.

A little known instability occurring within flux tubes is described by Marklund (1979) who suggested that if an electric field is present in a plasma permeated by a magnetic field and has a component perpendicular to the field, the {\bf E} x {\bf B} force will cause electrons and ions, but not neutral particles, to migrate to the axis of a flux tube.  As Peratt \& Verschuur (2010) note, because of different ionization potentials of various atomic species and cooling within the filaments, ionic species will then separate within a flux tube.  This mechanism, akin to the FIP effect, also has the seeds for separating the HI from the electrons.    

Another possibility for separating electrons and neutrals may involve interacting magnetically controlled filaments.  Fig. 1 shows two HI features, South Pair, which straddle a high-frequency continuum source, as displayed in Fig. 4a.  In that same direction the data indicate that a number of filaments of HI intersect, see Fig.1.  This raises the interesting possibility that magnetic reconnection may play a role in creating pockets of HI that are pulled away from a central X-neutral point where magnetic fields, likely to be present in the filaments, are reconnecting.  The continuum radiation revealed in the {\it ILC} data is then being produced at the X-neutral point.  Unfortunately the necessary theory to help account for the data invoking magnetic reconnection does not yet appear to exist.  Priest \& Forbes (2000) note that concerning the question of what actually happens to the particles at the X-point as regards energies and spectra, and how magnetic energy is converted to heat, kinetic energy and particle energy is largely unknown.  They state that $"$These apparently simple questions have not yet been answered fully [and] the answers are likely to be highly complex.$"$  In this context they list at least 12 possible types of reconnection that may play a role.  While their work also focused on events in the solar corona, future consideration of what may be occurring in interstellar space could clarify why so many associations exist between HI structures and high-frequency continuum peaks in the {\it ILC} data. 

These tentative suggestions should not cause us to avoid what the data show, that galactic HI peaks (clouds?) and {\it ILC} peaks are associated and slightly offset from one another, perhaps even along the axes of one or more of the filaments that pervade a volume of interstellar space.  The fact that even a simplistic consideration of the magnitudes of the parameters required to evaluate Eqtn. 1, Fig. 9, leads to reasonable values of the distance and electron densities is surely sufficient evidence that the issue is deserving of further study, given the importance attached to the conventional explanation of the {\it ILC} structure being at cosmological distances.

\section{Conclusions}
Galactic HI profiles in an area bounded by longitudes 85\arcdeg\ \& 110\arcdeg and latitudes 55\arcdeg and 65\arcdeg were decomposed into Gaussians.  The area was divided into two sections separated at l=85.¡5 and the analysis proceeded many months apart without cross-talk between them.  In all, 588 profiles at 0.\arcdeg5 spacing netted 3,706 Gaussian components.  These were sorted into 18 families, each defined by similarity of center velocity and line width.  When the average velocities of the derived Gaussian families are used to focus attention on specific HI channel maps, associations between HI peaks and  structure in the Internal Linear Combination ({\it ILC}) map of Hinshaw et al. (2007), based on the {\it WMAP} survey data, are dramatically revealed.  These associations are confirmed by comparing the morphology of the column densities of the Gaussian families and the {\it ILC} structures.  The step of first identifying Gaussian families in a given area of sky and using their center velocities as a guide, makes the task of finding associations with the {\it ILC} peaks manageable because only a fraction of the up to 100 channel maps that could be created for any area then need to be studied.  In addition, given that the Gaussian families summarize the properties of {\it all} the HI along the line-of-sight in the area, statistical arguments about whether or not the associations are due to chance becomes less relevant. 

The relationships between some of the families of HI Gaussian components show that despite their velocity differences they are physically related to one another and hence at the same distance. This phenomenon was also reported in Paper 1.  This would never have been discovered if it were not for attention drawn to the properties of the HI because of the associations with small-scale  {\it ILC} structure.

The Gaussian mapping reveals details in the HI morphology of several components at widely different velocities, from 0 km s$^{-1}$ to $-$109 km s$^{-1}$.  In an area labeled North Pair, two HI features straddle an {\it ILC} source that is located at the point of overlap of two filaments at velocities of order $-$30 and $-$13 km s$^{-1}$.  Similarly, in an area named South Pair, two HI peaks straddle an {\it ILC} peak and here the HI consists of three families of components at intermediate velocities around $-$36 km s$^{-1}$ with average line widths of 14.7, 7.0 \& 4.6 km s$^{-1}$ found where HI filaments at distinctly different velocities, around 0 km s$^{-1}$ \& $-$38 km s$^{-1}$ overlap.  

The previously hypothesized mechanism for producing the high-frequency continuum radiation from interacting HI features in interstellar space involving free-free emission from electrons (Verschuur, 2010) is re-examined in the light of the new data.  It is found to account for the existence of the small-scale {\it ILC} peaks if the sources are located from 30 to 100 pc from the Sun.  The pervasive presence of interstellar electrons is revealed in observations of pulsar dispersion measures and to fit the model the of necessity cold electrons have to be clumped on scales that are similar to those seen in the HI distribution with densities from 10 to 25\% of the immediately adjacent HI peaks.   Associated H$\alpha$ radiation at the location of the {\it ILC} peaks is not expected because the source of electrons does not require the localized ionization of HI as was hypothesized in Paper 2.  

In order to determine unequivocally whether or not the claimed associations are real, higher resolution observations are required.  For example, Planck data should be compared with high-resolution HI observations obtained with suitably large radio telescopes, provided attention is focussed on high-latitude regions where the confusion created by having too much HI in the beam is minimized.  In the meantime caution should be exercised in drawing far-reaching cosmological conclusions from the {\it ILC} data that may be compromised by the presence of intervening galactic sources of high-frequency continuum radiation.

Dr. Joan Schmelz is thanked for patiently hearing me out while I struggled to make sense of what is reported here. I also am grateful for discussions with Gary Hinshaw,  Adolf Witt, John Raymond, Mahboubeh Asgari-Targhi, and Michael Cervetti for a useful discussion on statistics

{}

\clearpage

\begin{deluxetable}{ccccc}									
\tablecolumns{5}									
\tablewidth{0pc}									
\tablecaption{Area 1 Component Families}									
\tablehead{\colhead{Name} & \colhead{{\# Gaussians}}   & \colhead{Ave. Center Vel.}    & \colhead{Ave Line width} & \colhead{Peak $N_H$}\\{} & {}  & {(km s$^{-1}$)} & {(km s$^{-1}$)} & {($10^{18} cm^{-2}$)} }									
\startdata									
(1)  &  (2)  & (3)  &  (4) &  (5)\\
Broad LV	&	273	&	 -5.0$\pm$3.1	&	33.3$\pm$2.0	&	92.4	\\
Broad IV	&	273	&	 -35.8$\pm$6.4	&	33.1$\pm$2.2  	&	59.7	\\
LV1	&	273	&	 -1.5$\pm$2.6	&	14.7$\pm$2.3	&	91.6	\\
LV2	&	169	&	 -0.2$\pm$2.1	&	5.8$\pm$2.3	&	57.9	\\
LV3	&	46	&	 -0.4$\pm$3.1  	&	4.1$\pm$2.1	&	22.6	\\
LV4	&	95	&	 -5.1$\pm$1.3	&	7.8$\pm$2.3	&	43.3	\\
LV5	&	32	&	 -12.2$\pm$2.7	&	9.2$\pm$3.8	&	33.1	\\
LV6	&	41	&	 +9.1$\pm$2.7	&	8.1$\pm$3.1	&	31.4	\\
LV11	&	44	&	 -27.3$\pm$2.7	&	10.7$\pm$4.2	&	38.4	\\
LV12	&	52	&	 -19.2$\pm$2.5	&	9.8$\pm$2.8	&	50.6	\\
LV13	&	48	&	 -19.1$\pm$3.3	&	5.1$\pm$1.9	&	31.7	\\
LV14	&	-	&	-	&	-	&	-	\\
IV-A	&	216	&	 -38.6$\pm$3.9	&	14.7$\pm$2.9	&	141.3	\\
IV-B	&	107	&	 -38.2$\pm$4.8	&	7.0$\pm$2.8	&	129.5	\\
IV-C	&	50	&	 -35.3$\pm$2.9	&	5.0$\pm$2.4	&	41.6	\\
IV-D	&	-	&	-	&	-	&	-	\\
IV-HV	&	24	&	 -73.6$\pm$5.4	&	16.0$\pm$4.4	&	24.1	\\
HV	&	-	&	-	&	-	&	-	\\
\enddata									
\end{deluxetable}									

\begin{deluxetable}{ccccc}									
\tablecolumns{5}									
\tablewidth{0pc}									
\tablecaption{Area 2 Component Families}									
\tablehead{\colhead{Name} & \colhead{{\# Gaussians}}   & \colhead{Ave. Center Vel.}    & \colhead{Ave Line width} & \colhead{Peak $N_H$}\\{} & {}  & {(km s$^{-1}$)} & {(km s$^{-1}$)} & {($10^{18} cm^{-2}$)} }									
\startdata									
(1)  &  (2)  & (3)  &  (4) &  (5)\\
Broad LV	&	315	&	 -5.2$\pm$3.9	&	32.5$\pm$2.4	&	58.3	\\
Broad IV	&	315	&	 -46.6$\pm$4.6	&	33.0$\pm$2.5  	&	61.3	\\
LV1	&	315	&	 -0.5$\pm$2.3	&	14.8$\pm$2.2	&	58.3	\\
LV2	&	74	&	 -2.7$\pm$2.4	&	6.3$\pm$3.0	&	20.9	\\
LV3	&	-	&	-	&	-	&	-	\\
LV4	&	76	&	 -7.5$\pm$1.6	&	7.6$\pm$2.9	&	38.4	\\
LV5	&	76	&	 -12.4$\pm$3.8	&	9.3$\pm$2.9	&	26.0	\\
LV6	&	40	&	 +5.4$\pm$3.4	&	8.7$\pm$4.2	&	29.2	\\
LV11	&	54	&	 -27.9$\pm$2.9	&	13.4$\pm$3.5	&	30.3	\\
LV12	&	55	&	 -21.7$\pm$3.1	&	9.8$\pm$3.3	&	40.8	\\
LV13	&	-	&	-	&	-	&	-	\\
LV14	&	48	&	 -25.7$\pm$5.6	&	3.6$\pm$1.4	&	31.1	\\
IV-A	&	281	&	 -48.0$\pm$4.9	&	14.7$\pm$3.0	&	48.7	\\
IV-B	&	145	&	 -45.7$\pm$5.4	&	7.4$\pm$2.4	&	24.6	\\
IV-C	&	24	&	 -38.2$\pm$6.2	&	6.6$\pm$1.9	&	10.3	\\
IV-D	&	40	&	 -54.6$\pm$3.5	&	9.7$\pm$3.8	&	25.3	\\
IV-HV	&	50	&	 -77.8$\pm$7.4	&	21.0$\pm$6.1	&	22.8	\\
HV	&	48	&	 -109.3$\pm$12.6	&	24.6$\pm$5.3	&	32.3	\\
Other	&	6	&	 -64.9$\pm$3.8	&	8.6$\pm$3.0	&	8.8	\\
\enddata									
\end{deluxetable}																					
\clearpage

\begin{deluxetable}{ccccc}									
\tablecolumns{5}									
\tablewidth{0pc}									
\tablecaption{Properties of Gaussian Families}									
\tablehead{\colhead{Name} & \colhead{Total $N_H$}   & \colhead{Fraction}    & \colhead{Ave $N_H$} & \colhead{Peak $N_H$}\\ {({\it l,b})} & {($10^{18} cm^{-2}$)}  & {of total} & {($10^{18} cm^{-2}$)} & {($10^{18} cm^{-2}$)} }									
\startdata									
(1)  &  (2)  & (3)  &  (4) &  (5)\\
Broad LV	&	14,749	&	0.204	&	25.1	&	92.4	\\
Broad IV	&	13,372	&	0.185	&	22.7	&	61.3	\\
LV1	&	18,759	&	0.259	&	31.9	&	91.6	\\
LV2	&	2,965	&	0.041	&	12.2	&	57.9	\\
LV3	&	356	&	0.005	&	7.7	&	22.6	\\
LV4	&	1,906	&	0.026	&	11.2	&	47.3	\\
LV5	&	856	&	0.012	&	7.4	&	33.1	\\
LV6	&	557	&	0.008	&	6.4	&	31.4	\\
LV11	&	1,031	&	0.014	&	10.3	&	38.4	\\
LV12	&	1,560	&	0.022	&	14.6	&	50.6	\\
LV13	&	410	&	0.006	&	9.8	&	31.7	\\
LV14	&	418	&	0.006	&	8.7	&	31.1	\\
IV-A	&	10,640	&	0.147	&	21.4	&	141.3	\\
IV-B	&	2,800	&	0.039	&	11.1	&	129.5	\\
IV-C	&	544	&	0.008	&	7.4	&	41.6	\\
IV-D	&	394	&	0.005	&	9.9	&	25.3	\\IV-HV	&	574	&	0.008	&	7.8	&	24.1	\\
HV	&	501	&	0.007	&	10.4	&	32.3	\\
-	&	-	&	-	&	-	&	-	\\
Grand total	&	72,392	&	-	&	-	&	-	\\
\enddata									
\end{deluxetable}

\clearpage

\begin{figure}
\figurenum{1a-b}
\epsscale{0.7}
\plotone{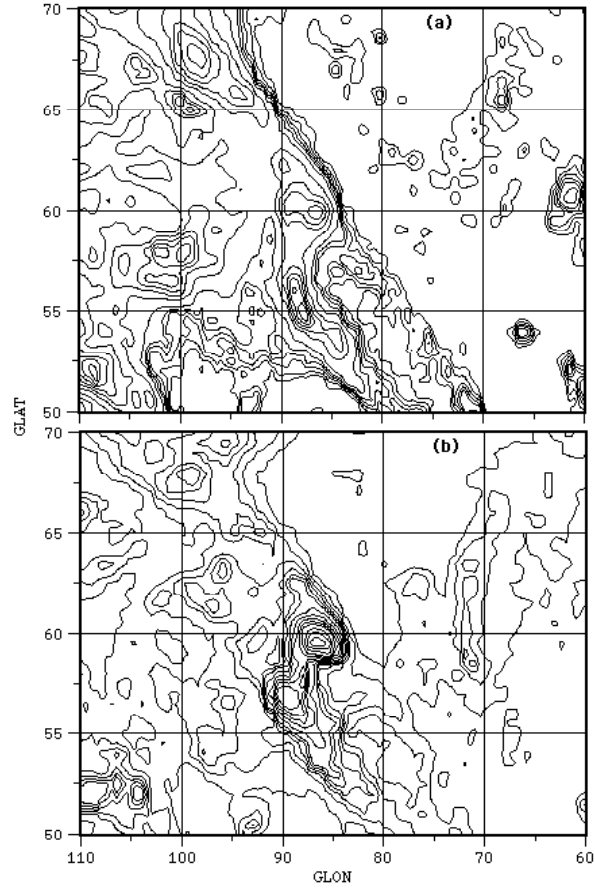}
\caption{Contour maps of the HI brightness in a 3.3 km s$^{-1}$ band centered at the following velocities.  (a) At $-$46 km s$^{-1}$, contours 0.5:1.5@0.2, 1.7:7.7@1 K.km s$^{-1}$.  The bright filamentary feature that curves from the upper left to the lower center is referred to as Filament Alpha and it shows weak parallel strand below {\it b} $=$57\arcdeg.  (b) At $-$36 km s$^{-1}$, contours 1:5@0.8, 7:20@3, 26:40@6 K.km s$^{-1}$.  The central pair of features is referred to as South Pair, see text.  }
\end{figure}

\clearpage

\begin{figure}
\figurenum{1c-d}
\epsscale{0.7}
\plotone{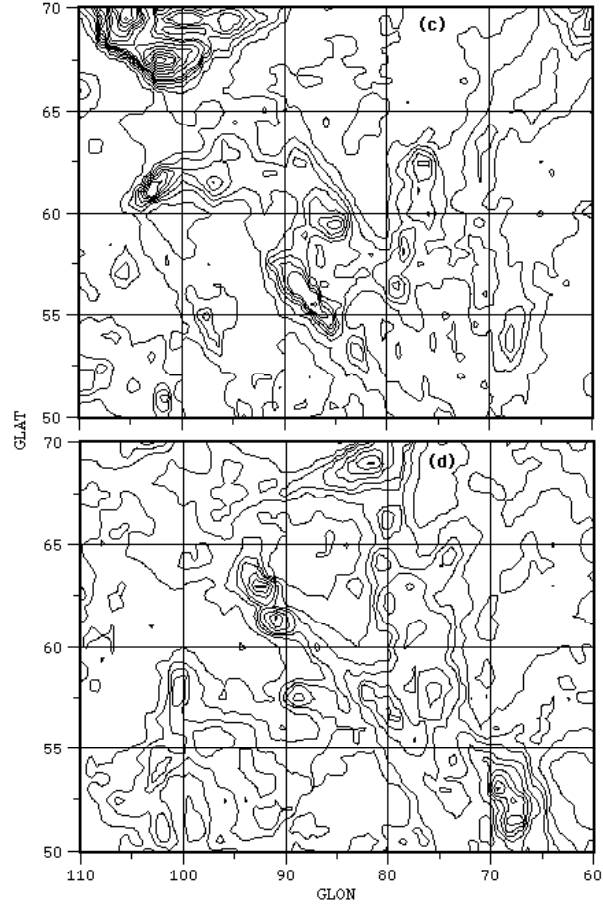}
\caption{(c)  At $-$30 km s$^{-1}$, contours 1:4@1, 7:28@3 K.km s$^{-1}$ The curved feature at the center is referred to as Filament Gamma in the text.  (d) At $-$13 km s$^{-1}$, contours 1:5@1, 7:17@2 K.km s$^{-1}$.  The diagonal feature is referred to as Filament Delta in the text.  It encompasses a pair of peaks around {\it l,b} $=$ (92\arcdeg, 62\arcdeg) called North Pair, see text.}
\end{figure}

\clearpage

\begin{figure}
\figurenum{1e-f}
\epsscale{0.7}
\plotone{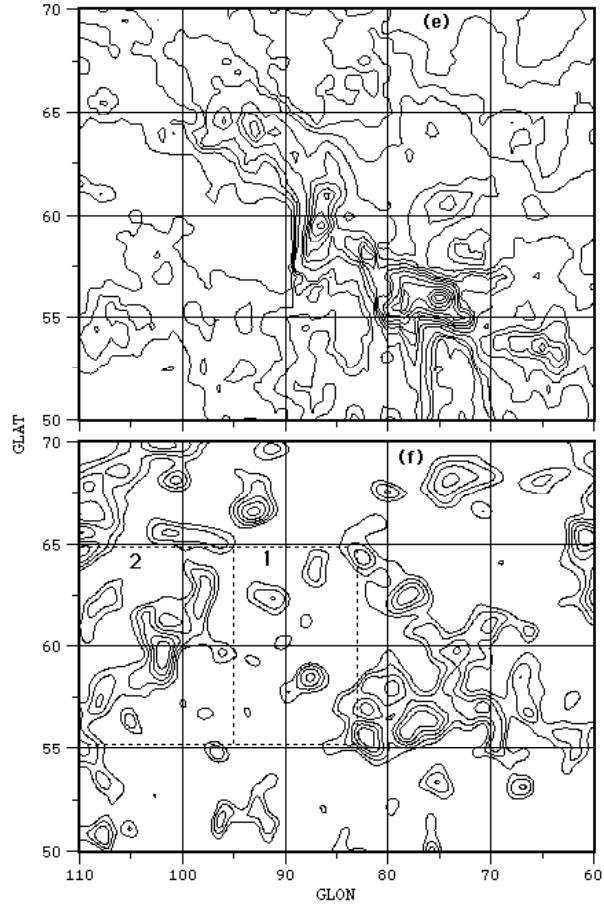}
\caption{(e)  At $+$1 km s$^{-1}$ in upper frame, contours 4:7@1, 10:22@3, 26:42@4 K.km s$^{-1}$.  The diagonal feature is referred to as Filament Beta in the text.  (f) The  {\it ILC} amplitudes for the area displayed as contours 0.02:0.17@0.03 mK. The two areas over which Gaussian analysis was performed are indicated by the dashed lines and the numerals 1 \& 2, see below.}
\end{figure}

\clearpage

\begin{figure}
\figurenum{2a-d}
\epsscale{1.0}
\plotone{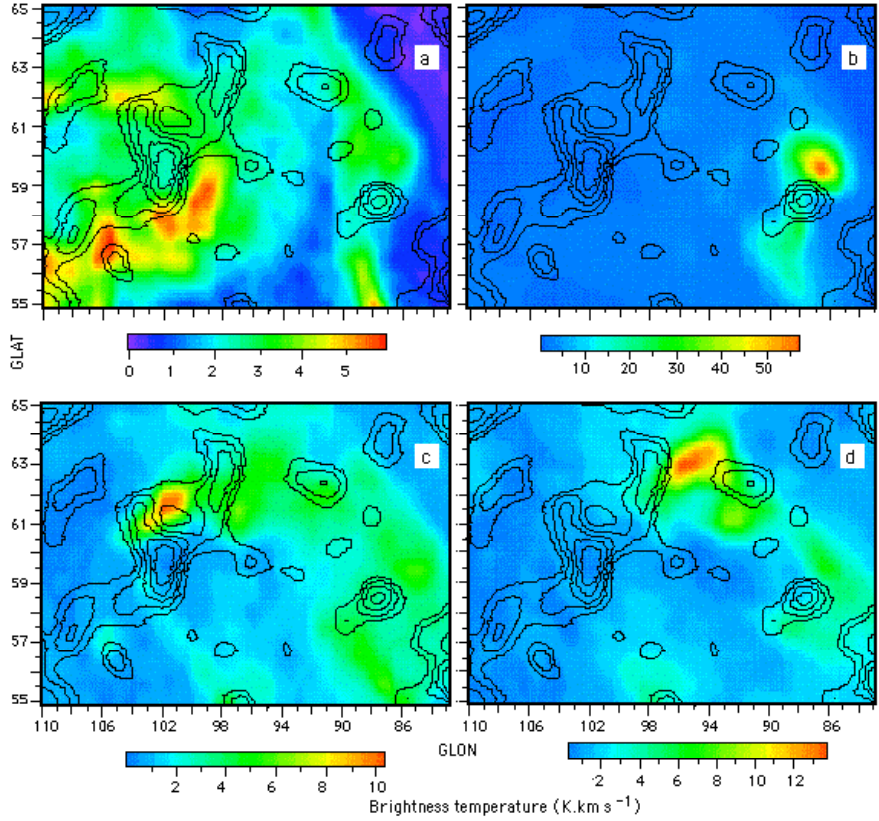}
\caption{Images of the HI brightness in a 3.3 km s$^{-1}$ band at velocities pertinent to the derived Gaussian families with the {\it ILC} contours overlain.  The HI brightnesses are indicated in the legends and the {\it ILC} contours are the same as for Fig. 1c.  (a) HI at $-$48 km s$^{-1}$. (b) HI at $-$38 km s$^{-1}$.  South Pair is seen here.  (c) HI at $-$28 km s$^{-1}$.   (d) HI at $-$20 km s$^{-1}$.  See text for a description.}
\end{figure}

\clearpage

\begin{figure}
\figurenum{2e-h}
\epsscale{1.0}
\plotone{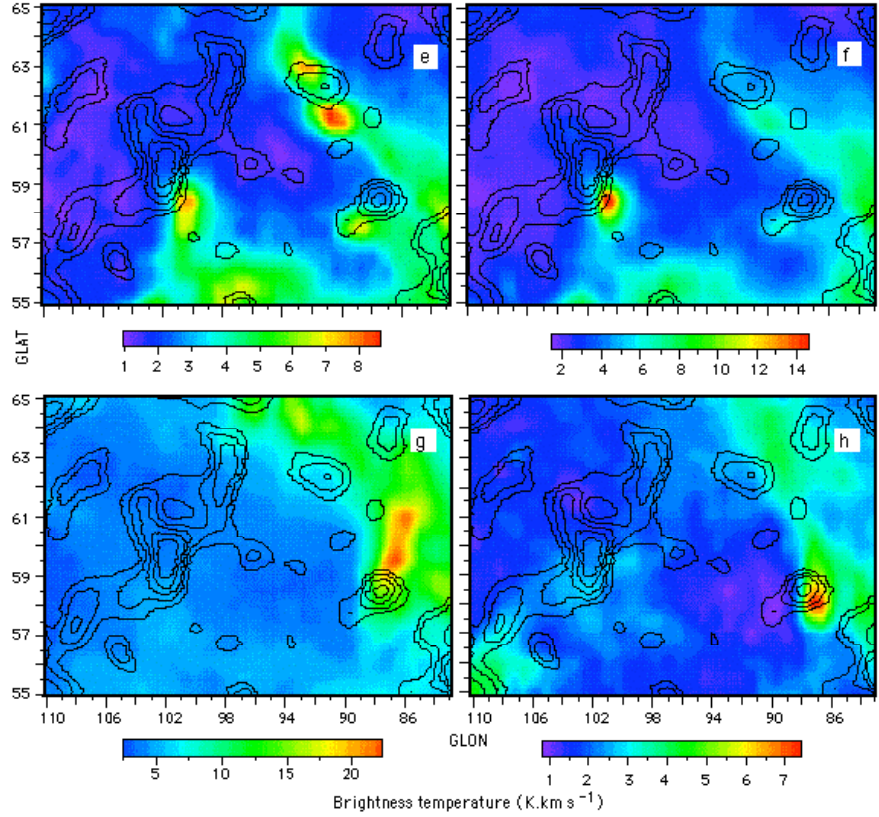}
\caption{(e) HI at $-$12 km s$^{-1}$. North Pair is at the upper right and the signature of S16, see text, is at  {\it l,b} $=$ (100\arcdeg, 58\arcdeg).  (f) HI at $-$9 km s$^{-1}$.   The southern half of North Pair is barely visible but S16 is dominant, see text.  (g) HI at 0 km s$^{-1}$.  (h) HI at $+$8 km s$^{-1}$.}
\end{figure}

\clearpage

\begin{figure}
\figurenum{3}
\epsscale{1.0}
\plotone{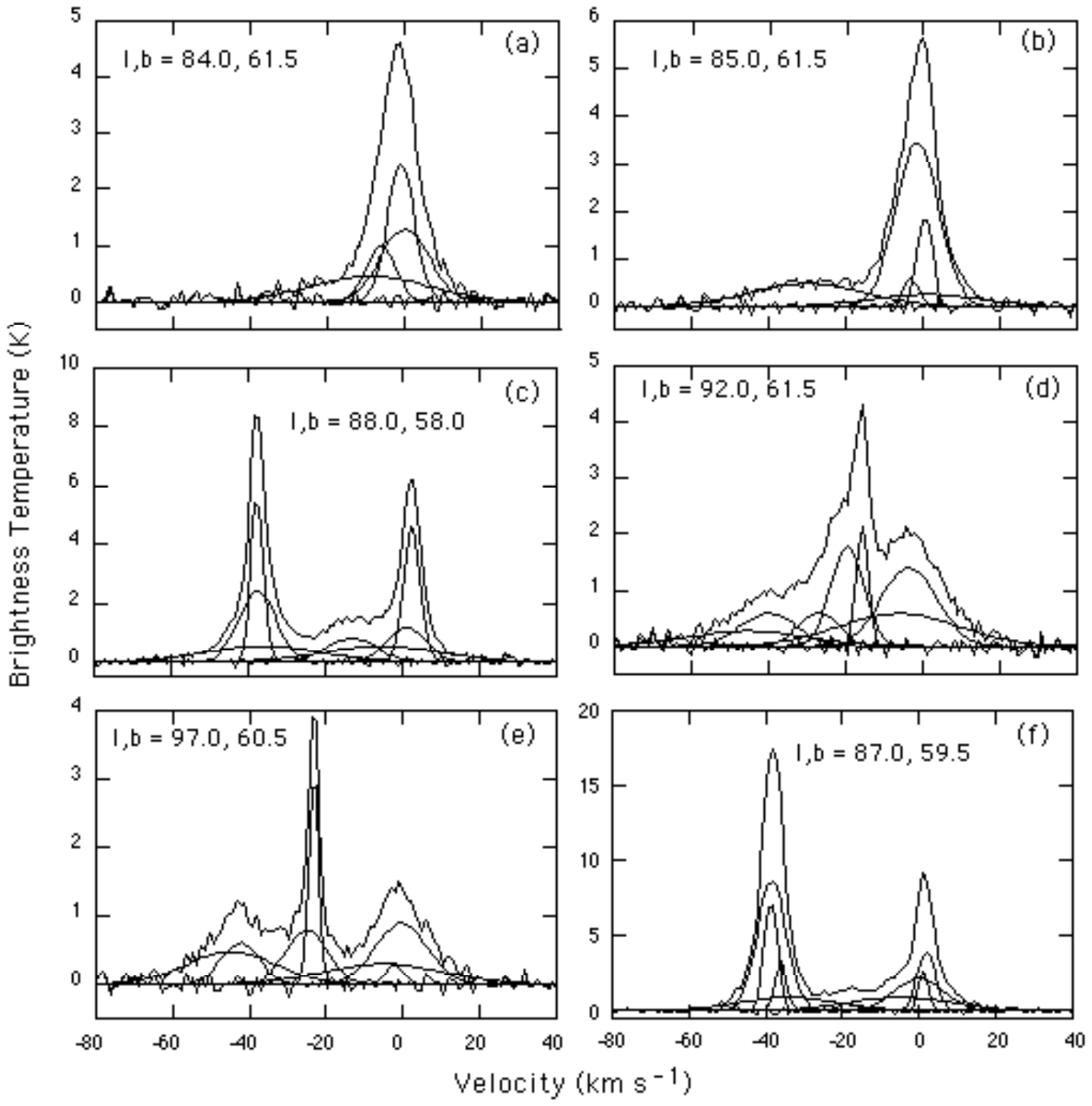}
\caption{Examples of Gauss decompositions discussed in the text for profiles in the directions indicated.}
\end{figure}

\clearpage

 \begin{figure}
\figurenum{4}
\epsscale{1.0}
\plotone{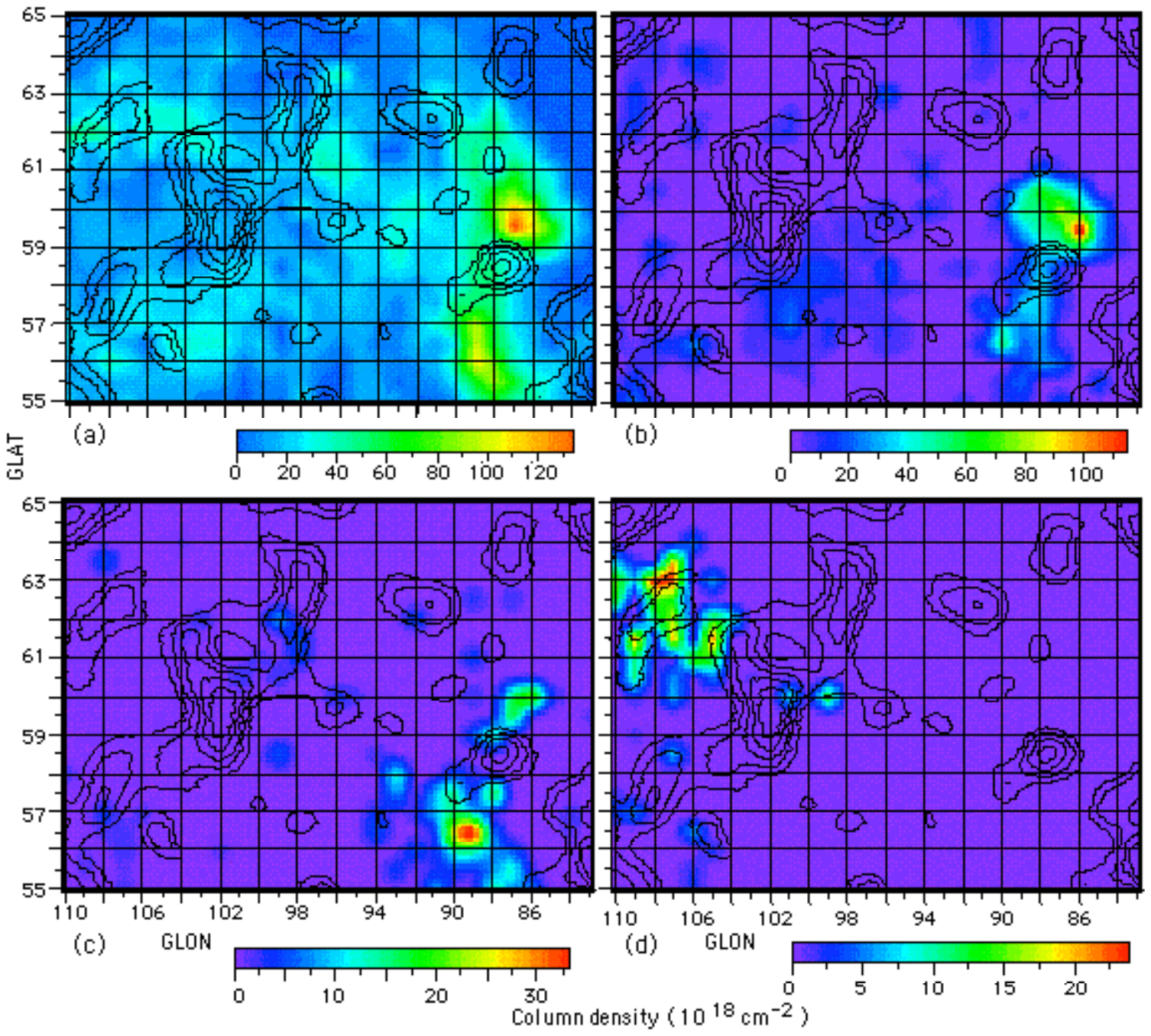}
\caption{The HI column density distributions for families of Gaussian components at intermediate velocities with the WMAP contours overlain, same contour levels as in Fig. 1c. (a) IV-A. (b) IV-B. (c) IV-C.  (d) IV-D.  See text. }
\end{figure}

\clearpage

\begin{figure}
\figurenum{5}
\epsscale{1.0}
\plotone{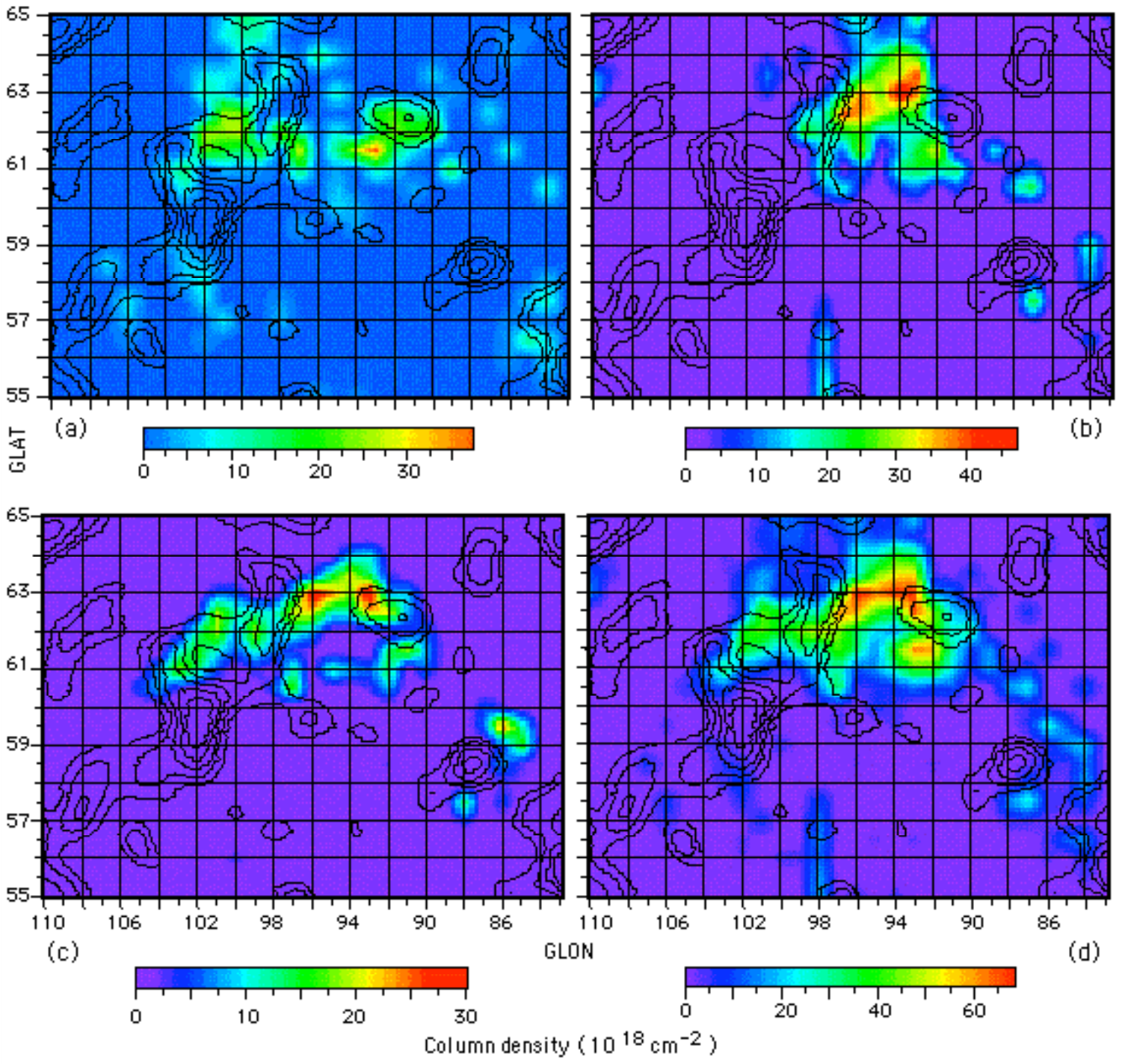}
\caption{The HI column density distributions for three families of Gaussian components at velocities between those included in figs. 5 \& 6, again with the WMAP contours overlain, same contour levels as in Fig. 1c. (a) LV11. (b) LV12. (c) LV13 \& 14 combined.  (d)  Sum of LV11, 12, 13 \& 14.  See text.}
\end{figure}

\clearpage

\begin{figure}
\figurenum{6}
\epsscale{0.9}
\plotone{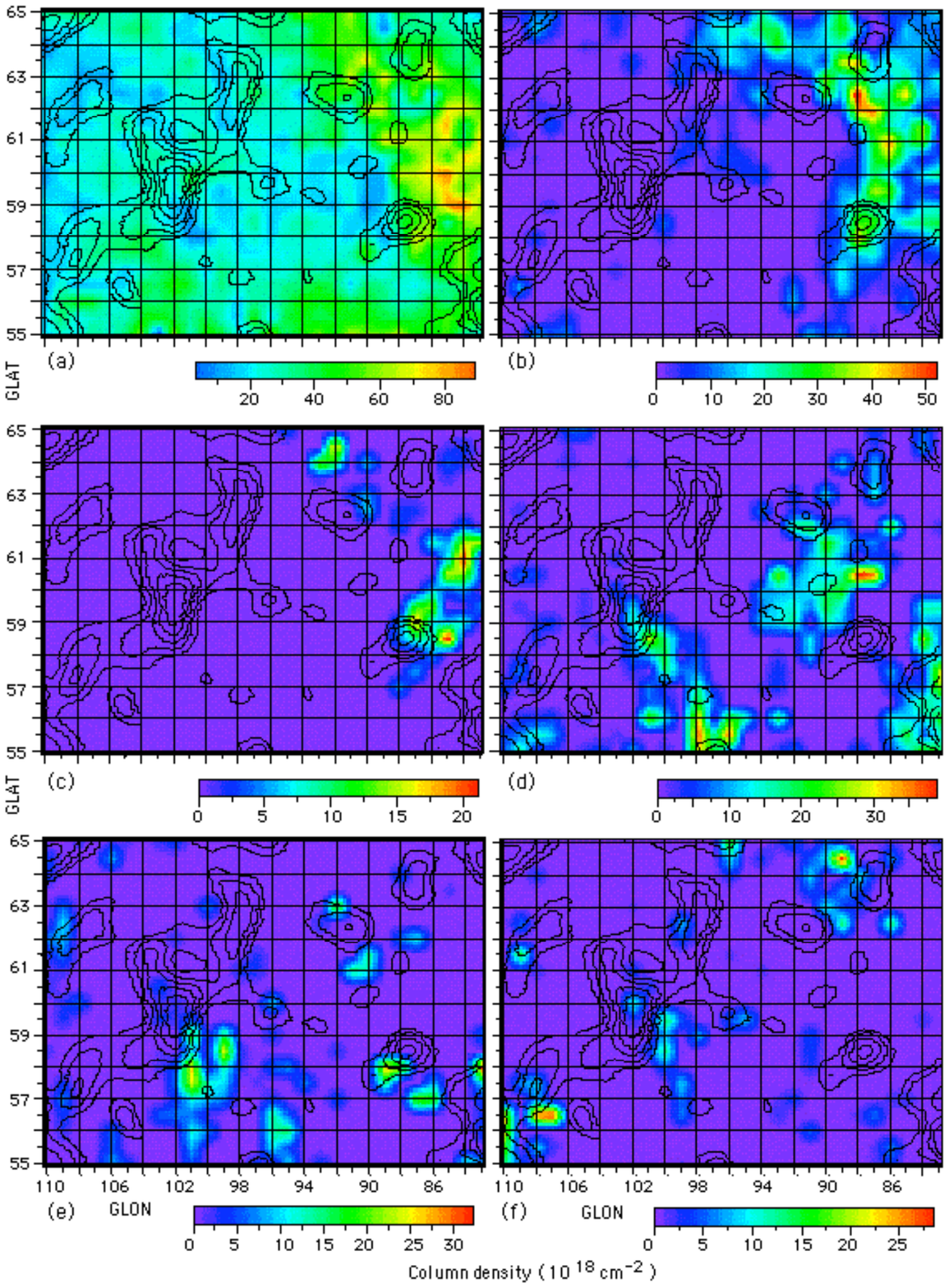}
\caption{The HI column density distributions for six families of Gaussian components at low velocities with the WMAP contours overlain, same contour levels as in Fig. 1c. (a) LV1, a pervasive component seen in all directions over the area. (b) LV2. (c) LV3.  (d) LV4. (e) LV5. (f) LV6.  See text.}
\end{figure}

\clearpage

\begin{figure}
\figurenum{7}
\epsscale{0.9}
\plotone{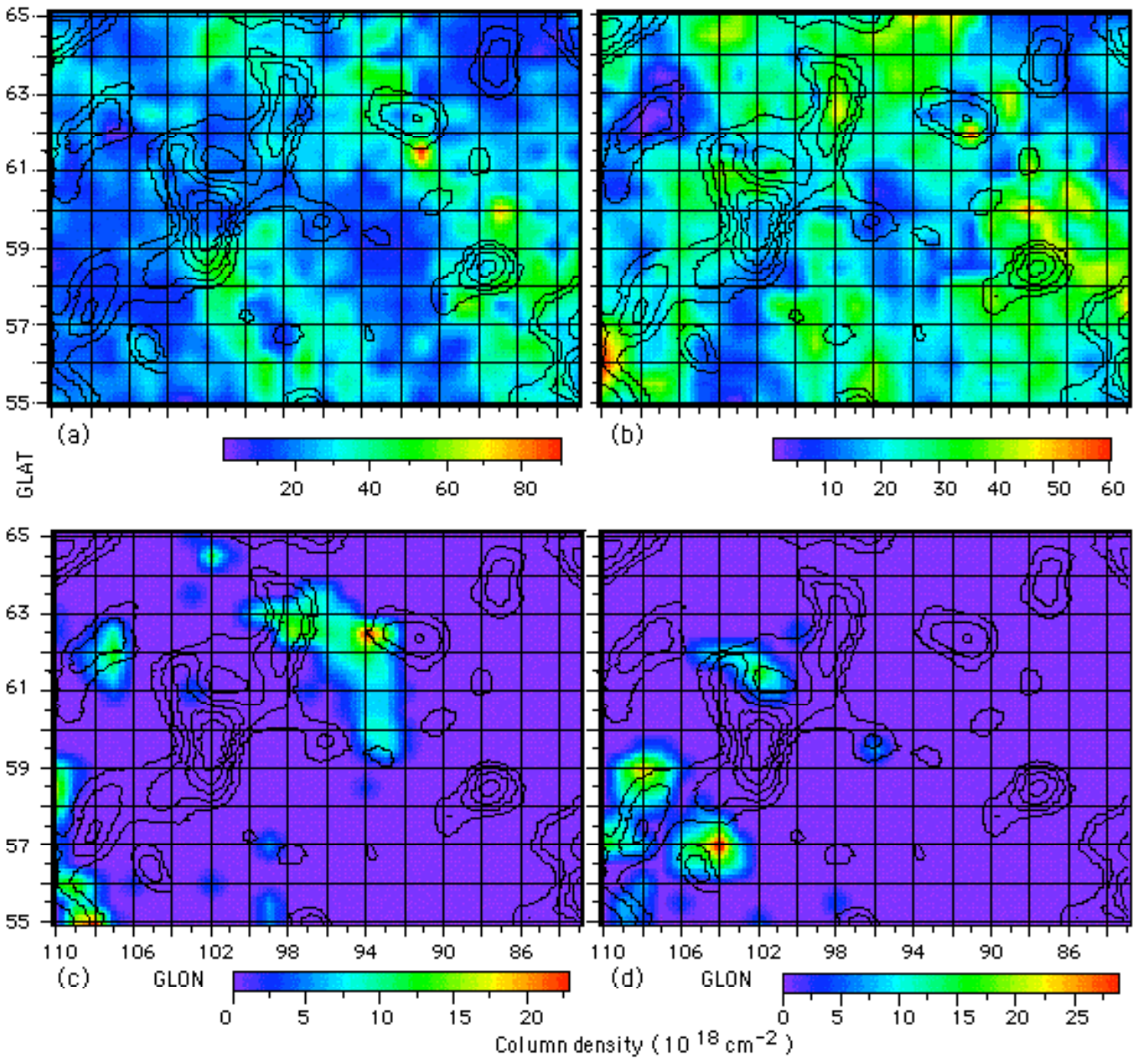}
\caption{The HI column density distributions for four remaining families of Gaussian components with the WMAP contours overlain, same contour levels as in Fig. 1c. (a) The low-velocity broad line width components, which are pervasive over the area. (b) Similar to a) but for intermediate velocity broad line width components that are also pervasive. (c) A limited number of components making up a family called IV-HV.  (d) High-velocity components HV.  See text.}
\end{figure}

\clearpage

\begin{figure}
\figurenum{8}
\epsscale{.7}
\plotone{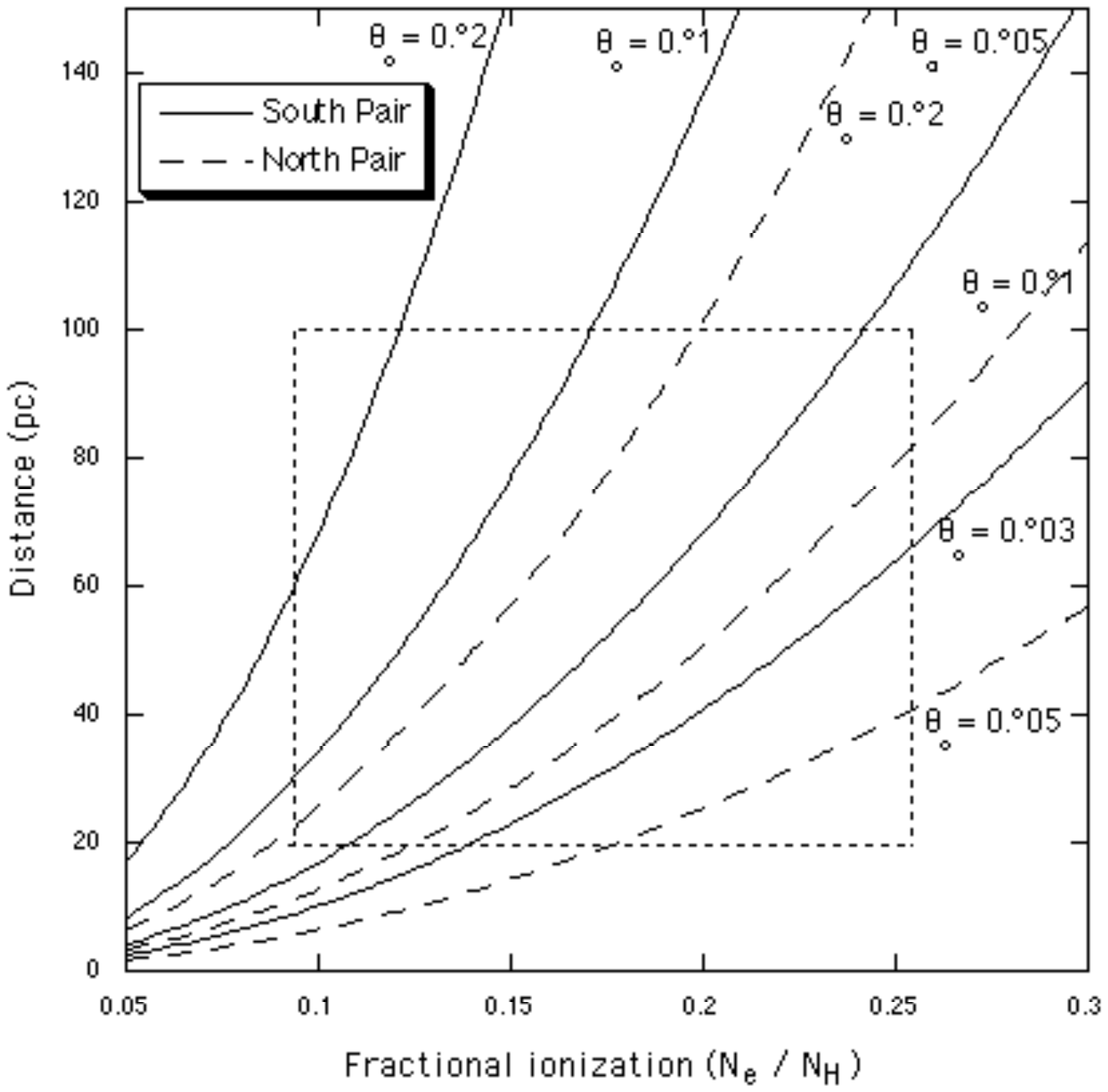}
\caption{The distance required to match the {\it ILC} amplitude for two cases indicated as a function of the electron column density expressed as a fraction of the associated HI peak.  The curves represent the values derived for different assumed angular widths of unresolved features.  The dashed line square is the suggested regime where the model calculations, \S7.2, give reasonable distances for reasonable values of the parameters required to evaluate Eqtn. 1, see text. }
\end{figure}

\clearpage

\end{document}